\documentclass[sigconf,screen]{acmart}
\usepackage{xspace}
\usepackage{longtable}
\usepackage{array}
\usepackage{multirow}
\usepackage{tabularx}

\AtBeginDocument{%
  }

\setcopyright{acmlicensed}
\copyrightyear{2018}
\acmYear{2018}
\acmDOI{XXXXXXX.XXXXXXX}
\acmBooktitle{Companion Proceedings of the 34th ACM Symposium on the Foundations of Software Engineering (FSE ’26), July 5 - 9, 2026 Montreal, Canada}

\acmISBN{978-1-4503-XXXX-X/2018/06}

\begin{document}

\title{Fairness Across Fields: Comparing Software Engineering and Human Sciences Perspectives}

\author{Lucas Valença}
\email{lucas.rodriguesvalen@ucalgary.ca}
\affiliation{%
  \institution{University of Calgary}
  \city{Calgary}
  \state{Alberta}
  \country{Canada}
}

\author{Ronnie de Souza Santos}
\email{ronnie.desouzasantos@ucalgary.ca}
\affiliation{%
  \institution{University of Calgary}
  \city{Calgary}
  \state{Alberta}
  \country{Canada}
  }

\begin{abstract}
\textbf{Background.} As digital technologies increasingly shape social domains such as healthcare, public safety, entertainment, and education, software engineering has engaged with ethical and political concerns primarily through the notion of algorithmic fairness. \textbf{Aim.} This study challenges the limits of software engineering approaches to fairness by analyzing how fairness is conceptualized in the human sciences. \textbf{Methodology.} We conducted two secondary studies, exploring 45 articles on algorithmic fairness in software engineering and 25 articles on fairness from the humanities, and compared their findings to assess cross-disciplinary insights for ethical technological development.
\textbf{Results.} The analysis shows that software engineering predominantly defines fairness through formal and statistical notions focused on outcome distribution, whereas the humanities emphasize historically situated perspectives grounded in structural inequalities and power relations, with differences also evident in associated social benefits, proposed practices, and identified challenges. \textbf{Conclusion.} Perspectives from the human sciences can meaningfully contribute to software engineering by promoting situated understandings of fairness that move beyond technical approaches and better account for the societal impacts of technologies.
\end{abstract}






\maketitle

\section{Introduction}
\label{sec:introduction}

Fairness has become a central concern in contemporary software engineering, particularly with the growing use of artificial intelligence systems that make or support decisions with direct social consequences \cite{brun2018software, soremekun2022software}. These systems are increasingly deployed in domains such as healthcare, hiring, education, and finance, where their outputs shape opportunities, access, and life outcomes \cite{brun2018software, ferrara2024fairness, soremekun2022software, lecca2025towards}. As a result, expectations of software quality have expanded beyond technical correctness to include ethical accountability and social responsibility \cite{de2025software, brun2018software}. In this context, fairness is often treated as a quality attribute that seeks to ensure equitable system behavior across social groups, aiming to prevent systematic advantage or disadvantage associated with characteristics such as gender, race, or socioeconomic status \cite{verma2018fairness, ferrara2024fairness, de2025software}.

Within software engineering, fairness is commonly operationalized through computational definitions that can be measured, tested, and verified \cite{verma2018fairness, soremekun2022software}. This orientation has led to a proliferation of formal notions such as group, individual, and causal fairness, alongside metrics, testing strategies, and bias mitigation techniques designed to assess and correct algorithmic behavior \cite{verma2018fairness, ferrara2024fairness, brun2018software, soremekun2022software, chen2024fairness}. While these approaches provide actionable mechanisms for engineering practice, they tend to frame fairness as a technical property of data, models, or systems. In doing so, they often abstract away the social relations, institutional contexts, and power dynamics within which AI systems are designed, deployed, and interpreted \cite{de2025software}. As a consequence, fairness mechanisms may define human judgment and decision-making in software engineering without fully accounting for their broader socio-technical implications.

In contrast, the human and social sciences have long explored fairness as a core ethical and social concept \cite{rawls1958justice, dator2006fairness, hayibor2017fair, baumann2023fairness, varona2022discrimination}. Disciplines such as sociology, law, philosophy, and communication conceptualize fairness as context-dependent, culturally situated, and closely tied to questions of justice, participation, and legitimacy \cite{varona2022discrimination, baumann2023fairness, hayibor2017fair}. From these perspectives, fairness cannot be reduced to a single metric or formal property, but emerges through social processes, institutional arrangements, and collective deliberation. The contrast between these perspectives highlights a disciplinary gap: while software engineering emphasizes operationalization and verification, the human sciences emphasize interpretation, normativity, and lived experience \cite{brun2018software, ferrara2024fairness, de2025software, lecca2025towards, carey2022causal, staron2024laws}. Bridging this gap is important for understanding how AI-based fairness mechanisms mediate human reasoning and responsibility in software engineering practice.

This study addresses this challenge by comparing and contrasting how fairness is defined and approached in software engineering and in the human and social sciences. Specifically, we investigate the following research question: {\textit{\textbf{RQ.} How does the concept of fairness in software engineering differ from its interpretation in the human and social sciences?}} To answer this question, we conducted a comparative analysis that integrates a systematic tertiary review of software engineering literature with a rapid review of work from the human sciences.  Our findings indicate that software engineering research predominantly frames fairness as a measurable and testable property associated with bias mitigation and algorithmic evaluation, whereas the human sciences conceptualize fairness as a multidimensional social value encompassing justice, inclusion, and participation. 

By synthesizing and contrasting these perspectives, this study contributes to human-centered AI research in software engineering by clarifying how technical fairness mechanisms interact with human judgment and social values, and by identifying conceptual limitations that arise when fairness is treated solely as a computational attribute. The study further offers directions for aligning software engineering approaches with broader socio-ethical understandings of fairness, supporting more reflective and responsible AI-mediated software practices.


\section{Background}
\label{sec:background}

Software fairness has emerged as a non-functional requirement in contemporary software engineering, gaining importance with the rise of artificial intelligence and machine learning systems. Fairness is increasingly treated as a property that can be defined, measured, and verified alongside performance, reliability, or security \cite{verma2018fairness, soremekun2022software, chen2024fairness}. Foundational work has organized fairness definitions into families such as group, individual, and causal fairness, each describing how systems should behave equitably toward different users or social groups \cite{verma2018fairness}. These formalizations enable fairness to be operationalized through quantitative methods that detect and mitigate disparities in algorithmic outcomes, leading to its conceptualization as a measurable quality attribute integrated into testing and verification activities \cite{chen2024fairness, soremekun2022software}. However, such approaches often abstract fairness from its ethical and social underpinnings, reducing complex moral questions to computational criteria. In response, scholars have described fairness as a socio-technical construct linking technical design decisions to social accountability, introducing the notion of fairness debt to capture the ethical consequences of biased data and design practices that accumulate over time \cite{de2025software}. Despite this growing awareness, empirical studies indicate that fairness remains a secondary quality goal in practice, often overshadowed by accuracy and performance priorities \cite{ferrara2024fairness}.

Research on the human and organizational dimensions of software engineering further shows that fairness concerns extend beyond algorithms and datasets. Engineers frequently discuss fairness in relation to recruitment, compensation, collaboration, and representation, emphasizing that it is embedded in the social structures of software work \cite{sesari2024understanding}. These perspectives align with calls to situate fairness within broader ethical frameworks encompassing transparency, responsibility, and respect for diversity \cite{staron2024laws}. Interdisciplinary studies highlight that fairness is shaped by the legal, cultural, and institutional contexts of software development, indicating that equitable systems depend as much on social organization as on algorithmic design \cite{ferrara2024fairness, soremekun2022software, de2025software}. Overall, fairness in software engineering emerges as a dynamic and interpretive attribute formed through the interaction of technical mechanisms, human judgment, and societal structures, requiring both quantitative verification and qualitative reflection to be meaningfully realized.
\section{Method}
\label{sec:method}

This research employed two complementary review methodologies to synthesize fairness research across disciplines. A systematic tertiary review in software engineering was used to map how fairness is defined, operationalized, and evaluated, focusing exclusively on existing systematic reviews following established guidance \cite{kitchenham_systematic_2010}. This approach is appropriate for software engineering, where the growing volume of systematic reviews on fairness, artificial intelligence, machine learning, and responsible software development has produced a substantial body of secondary evidence, enabling a broad synthesis of shared definitions, recurring metrics, and methodological gaps. In parallel, a rapid review was conducted in the humanities and social sciences to synthesize evidence on algorithmic fairness using systematic yet simplified procedures suited to fast moving research areas \cite{cartaxo_rapid_2020}. This method supports timely capture of influential theoretical and empirical discussions in fields where fairness debates evolve quickly in response to technological and social developments, enabling cross disciplinary comparison between technical and sociotechnical perspectives.

\subsection{Research Questions}

We defined the same research questions for both systematic reviews to ensure conceptual alignment and comparability between the engineering and humanities analyses. The study addresses the following questions: {\textbf{RQ1.}} \textit{How is fairness defined and conceptualized?} {\textbf{RQ2.}} \textit{What metrics and criteria are used to evaluate fairness?} {\textbf{RQ3.}} \textit{What social benefits are associated with fair practices, processes, and technologies?} {\textbf{RQ4.}} \textit{What approaches, methods, and practices are proposed to promote fairness in technologies?} {\textbf{RQ5.}} \textit{What limitations and challenges are identified in the efforts to achieve fairness in technologies?} However, during the humanities rapid review, it became evident that {\textbf{RQ2}} (\textit{What metrics and criteria are used to evaluate fairness?}) was not applicable, as the notion of metrics pertains primarily to technical and engineering frameworks rather than the more conceptual analyses typical of the human and social sciences.

\subsection{Search Strategy}
To conduct the combined systematic reviews, we developed two complementary search strategies targeting different disciplinary scopes, while sharing a common conceptual focus on algorithmic fairness. The first review focused on fairness within the software engineering field, mapping definitions, evaluation criteria, approaches and limitations. For this review, we used conducted both manual and automatic searches in Google Scholar, ACM Digital Library, IEEE Xplore, Scopus, ScienceDirect, and Sage Journals, covering studies published since 2017. The search string used was:

\begin{quotation}
\footnotesize
    (“software fairness” OR “fairness in AI” OR “fairness in ML” OR “algorithmic discrimination” OR “algorithmic bias” OR “software discrimination” OR “software bias” OR “fairness in software engineering”) AND (“systematic review” OR “systematic literature review” OR “literature review” OR “survey” OR “mapping study”) AND (“software engineering”).
\end{quotation}

The second review focused on human and social science perspectives on algorithmic fairness, aiming to identify how fairness is conceptualized beyond engineering domains. An automatic search was conducted in Google Scholar, and no specific timeline was defined, as we focused on relevance to the theme, as suggested in the guidelines for rapid reviews. The following string guided the search:

\begin{quotation}
\footnotesize
("fairness" OR "justice" OR "ethics" OR "inequality" OR "discrimination" OR "algorithmic discrimination" OR "injustice" OR "social impact*" OR "societal impact*" OR "algorithmic harm*" OR "algorithmic inequality" OR "unfairness") AND ("algorithm*" OR "artificial intelligence" OR "AI" OR "machine learning" OR "ML" OR "generative AI" OR "AI system*" OR “software”) AND ("social science*" OR "sociology" OR "education" OR "law" OR "political science" OR "anthropology" OR "economics" OR "psychology" OR "communication" OR "media studies" OR "public policy" OR "science, technology and society" OR "STS" OR "sociotechnical" OR "critical data studies" OR "social theory")
\end{quotation}

\subsection{Selection Process and Data Extraction}
We established shared exclusion criteria and adapted inclusion criteria to the specific objectives and contexts of each study. Both inclusion and exclusion criteria are presented in Tables \ref{tab:criteria_combined}. After applying these criteria, a total of 70 studies were selected, with 45 from software engineering and 25 from the humanities. The software engineering studies were published between 2017 and 2025, while the studies from the human and social sciences that met our inclusion and exclusion criteria ranged from 2016 to 2025. For each included study, we developed structured extraction tables to systematically capture relevant information used to answer our research questions. The extracted data included core concepts and definitions of fairness, associated characteristics and dimensions, examples of applications and methods, and contextual details illustrating how fairness was operationalized and evaluated across domains.

\begin{table}[ht]
\scriptsize
\centering
\caption{Inclusion and exclusion criteria for both reviews (software engineering and humanities).}
\label{tab:criteria_combined}
\begin{tabularx}{\columnwidth}{|l|X|}
\hline
\textbf{ID} & \textbf{Description} \\
\hline
\multicolumn{2}{|l|}{\textbf{Inclusion Criteria (IC)}} \\
\hline
\multicolumn{2}{|l|}{\textit{Software Engineering Review}} \\
\hline
IC1 & The study is a literature review that addresses definitions of fairness in software engineering, including, but not limited to, factors that ground these definitions, how these definitions relate to algorithmic discrimination, and the criteria used to evaluate fairness. \\
\hline
\multicolumn{2}{|l|}{\textit{Humanities Rapid Review}} \\
\hline
IC1 & The study explores social impacts and dynamics of algorithms through the lenses of the human sciences (e.g., communication, sociology, anthropology, law), including, but not limited to, algorithmic discrimination, fairness, big data, and other sociotechnical dynamics enacted by algorithms and technologies. \\
IC2 & The study develops critiques of approaches for fair and ethical technologies and algorithms proposed by engineers. \\
\hline
\multicolumn{2}{|l|}{\textbf{Exclusion Criteria (EC)}} \\
\hline
\multicolumn{2}{|l|}{\textit{Software Engineering Review}} \\
\hline
EC1 & The study does not conduct a literature review of fairness or fairness testing. \\
EC2 & The study cannot be downloaded through the University of Calgary proxy. \\
EC3 & The study is not conducted in English. \\
EC4 & Duplicated studies: if one or more works have duplicates, only the most recent ones are considered. If a conference paper and its journal extension are selected, only the latter is included. \\
EC5 & The study is not conducted in the software engineering field. \\
EC6 & The study was published before 2017. \\
\hline
\multicolumn{2}{|l|}{\textit{Humanities Rapid Review}} \\
\hline
EC1 & The study does not explore social impacts of algorithms and technologies. \\
EC2 & The study cannot be downloaded through the University of Calgary proxy. \\
EC3 & The study is not conducted in English. \\
EC4 & Duplicated studies: if one or more works have duplicates, only the most recent ones are considered. If a conference paper and its journal extension are selected, only the latter is included. \\
EC5 & The study is not conducted in human science fields, including, but not limited to, communication, sociology, anthropology, and law. \\
\hline
\end{tabularx}
\end{table}

\subsection{Data Analysis and Synthesis}
The findings were synthesized using complementary quantitative and qualitative approaches. Descriptive statistics provided an overview of publication patterns, disciplinary focus, and temporal trends across the selected studies \cite{george2018descriptive}. These summaries helped identify how fairness research has evolved within software engineering and the humanities. Thematic synthesis was then applied to interpret how fairness was defined, measured, and theorized in each field \cite{cruzes2011recommended}. Codes derived from the studies were grouped into categories and refined into analytical themes, revealing convergences and divergences between technical and social interpretations of fairness. These analyses offered both scope and depth, linking empirical patterns with conceptual insights.

\subsection{Threats to Validity}
This study is subject to common validity threats associated with systematic reviews, including potential bias in study selection, database coverage, and researcher interpretation. To mitigate these risks, multiple databases were searched across both domains and cross-verified with reference lists to ensure completeness. Thematic synthesis followed established guidelines for qualitative analysis in software engineering \cite{cruzes2011recommended}, with one reviewer conducting the initial coding and another reviewing and refining it through biweekly one-hour discussion sessions until no disagreements remained. The integration of quantitative and qualitative synthesis provided methodological triangulation, strengthening reliability, although variations in reporting quality may still limit the comprehensiveness of the evidence.
\section{Findings} \label{sec:findings}

In this section, we present the results from the systematic reviews conducted to understand fairness in the fields of software engineering and human sciences.

\subsection{Software Engineering Tertiary Literature Review}
Our tertiary review analyzed 45 software engineering papers and shows that fairness research is largely centered on AI and ML decision-making systems and on formal and statistical fairness metrics, which account for approximately half of the studies. Bias and algorithmic discrimination appear frequently as core topics, with bias often treated as a condition that does not necessarily lead to discrimination. Fairness is commonly discussed alongside accountability and transparency within responsible AI and science, technology, and society perspectives, even when discrimination is not the primary focus. Only a small number of studies address fairness testing, application-specific contexts, or critical perspectives on prevailing fairness practices. Below, we present the evidence collected from the identified papers to answer the research questions.

\subsubsection{How is fairness defined and conceptualized?}
\label{sec:se_rq1}

The most common definition of fairness, which was present in the 45 analyzed texts, consists in the \textit{absence of discrimination through the distribution of results}. Distribution of results, in the context of digital systems, is tied to decision-making systems. Those systems have become very common due to the development of AI and ML. In other words, fairness is often understood as the absence of discrimination in the final decisions of AI and ML systems. Those decision-making processes are a result of several classifications (e.g., sharing advertising based on a person's data, such as gender and country). More than that, the result is itself a classification (e.g., classifying a person as guilty or non-guilty through a trial-assessment tool). Fairness studies add one more layer to this: they focus on classifying if a system of classifications is fair or not. The definition of \textit{fairness as a formal, mathematical and statistical property of the software} was found in 19 researches, this definition is justified because algorithms ``speak mathematics", so it becomes important to mathematically ``define aspects of society’s fundamentally vague notions of fairness and justice" \cite{rabonato_systematic_2025}. Continuing this formal comprehension of fairness, \cite{pham_fairness_2023} argue that fairness is an abstract notion that needs to be made concrete. The authors argue that AI and machine learning systems can only be considered trustworthy if their fairness can be verified and validated. They emphasize the need for a systematic approach to define fairness within the scope of the developing system, describe it as a software attribute that can be integrated into later phases of the development process, and ensure it can be tracked and verified throughout both development and operation.

\begin{table}[!h]
\centering
\scriptsize
\caption{Software engineering articles per fairness definition}
\label{tab:se_rq1}
\begin{tabular}{|p{0.5\columnwidth}|p{0.4\columnwidth}|}
\hline
\textbf{Definition of fairness} & \textbf{Articles} \\
\hline
Fairness concerns the absence of bias/discrimination through distribution of results — outcomes should not favor or harm individuals or groups unjustly. & P01--P45 \\
\hline
Fairness is defined by how decisions are made — the process must be impartial, consistent, and respectful. & P01, P02, P05, P12, P14, P15, P16, P17, P22, P23, P26, P27, P30, P34, P38, P39, P40, P43, P45 \\
\hline
Fairness is a formal property of the system. & P03, P04, P05, P08, P09, P10, P11, P13, P17, P20, P24, P25, P26, P29, P32, P37, P40, P41, P45 \\
\hline
\end{tabular}
\end{table}

\subsubsection{What metrics and criteria are used to evaluate fairness?} 
\label{sec:se_rq2}
The definitions presented in section \ref{sec:se_rq1} underpin the premise that fairness must be measurable and testable, which leads software engineers to define criteria to evaluate whether a system is fair. These criteria are grounded in binary classifier evaluation metrics, such as true positive, positive predicted value, true positive rate, and false positive rate. From these criteria, several fairness metrics emerge. The goal of this research is not to discuss metrics individually, but rather to understand what these sets of metrics represent. In other words, instead of asking “what metrics do engineers create?”, we aim to understand “what do the criteria that underlie fairness metrics mean?”. In this sense, fairness metrics reflect four primordial meanings: \textit{fair treatment of individuals}, \textit{fair treatment of groups}, \textit{fairness through awareness}, and \textit{fairness through unawareness}. The first two determine whether criteria focus on individuals or groups, while \textit{fairness through unawareness} assumes that ignoring sensitive attributes will prevent discrimination and \textit{fairness through awareness} explicitly engages with such attributes to achieve fair outcomes. These baselines are closely linked to the concepts of fair process and fair distribution of results. The most common metrics include equalized odds, demographic parity, equal opportunity, predictive equality, predictive parity, and counterfactual fairness, which impose constraints on outcome distributions across groups. Counterfactual fairness further requires that an individual’s prediction remain unchanged if a protected attribute were hypothetically altered. Overall, these definitions show that fairness metrics focus on the distribution of results and statistical parity between groups, without necessarily addressing the underlying causes of discrimination.

\begin{table}[!h]
\centering
\scriptsize
\caption{Software engineering articles per fairness metric}
\label{tab:se_rq2}
\begin{tabularx}{\columnwidth}{|X|X|}
\hline
\textbf{Fairness metric} & \textbf{Articles} \\
\hline
Group Fairness & P01--P45 \\
\hline
Individual Fairness & P01, P02, P03, P04, P05, P06, P07, P08, P09, P10, P11, P13, P14, P15, P16, P17, P20, P21, P22, P24, P25, P26, P27, P28, P29, P32, P33, P34, P35, P36, P37, P38, P39, P40, P41, P42, P43, P45 \\
\hline
Fairness through Awareness & P01, P03, P04, P05, P06, P07, P08, P09, P11, P12, P13, P14, P16, P21, P22, P23, P24, P25, P26, P27, P28, P29, P32, P34, P36, P37, P38, P39, P40, P41, P42, P43, P45 \\
\hline
Demographic Parity & P01, P03, P04, P05, P06, P07, P08, P09, P11, P13, P14, P16, P21, P22, P23, P24, P25, P26, P27, P31, P32, P34, P35, P36, P37, P38, P40, P41, P42, P43, P45 \\
\hline
Equalized Odds & P01, P04, P05, P06, P07, P08, P09, P11, P13, P14, P16, P21, P23, P24, P25, P26, P27, P31, P32, P34, P35, P36, P37, P38, P41, P42, P45 \\
\hline
Equal Opportunity & P01, P04, P05, P06, P07, P08, P09, P11, P14, P16, P21, P22, P24, P25, P26, P27, P31, P32, P34, P35, P36, P37, P38, P41, P42, P43 \\
\hline
Fairness through Unawareness & P01, P03, P04, P05, P06, P07, P08, P11, P13, P14, P16, P17, P21, P24, P25, P27, P32, P34, P35, P36, P37, P42, P45 \\
\hline
Causal or Counterfactual Fairness & P01, P03, P04, P05, P08, P09, P11, P12, P13, P14, P16, P21, P24, P25, P34, P36, P37, P40, P41, P43, P45 \\
\hline
Predictive Equality & P03, P07, P09, P11, P12, P13, P14, P16, P24, P27, P38, P40, P41, P42 \\
\hline
Predictive Parity & P03, P09, P11, P12, P14, P16, P24, P27, P38, P40, P41, P42 \\
\hline
Conditional Demographic Parity & P03, P05, P07, P11, P13, P14, P16, P21, P24, P36, P37 \\
\hline
Disparate Impact & P09, P12, P16, P22, P27, P31, P32, P33, P34, P35, P42 \\
\hline
Calibration & P03, P05, P14, P16, P21, P29, P41, P42 \\
\hline
Treatment Equality & P03, P06, P07, P11, P14, P16, P21, P24 \\
\hline
Intersectional Fairness & P01, P08, P11, P14, P16, P29, P41\\
\hline
No Proxy Discrimination & P01, P03, P04, P07, P14, P16 \\
\hline
Balance for Negative/Positive Class & P03, P07, P11, P14, P16 \\
\hline
No Unresolved Discrimination & P03, P04, P07, P14, P16 \\
\hline
Well Calibration & P03, P07, P14, P16, P40 \\
\hline
Overall Accuracy Equality & P03, P07, P14, P16 \\
\hline
Generalized Entropy Index & P07, P09, P22 \\
\hline
Preference-based Fairness & P24, P45 \\
\hline
Accuracy Rate Difference & P26 \\
\hline
Bounded Group Loss & P26 \\
\hline
Differential Fairness & P29 \\
\hline
Informational Fairness & P45 \\
\hline
Interactional Fairness & P45 \\
\hline
\end{tabularx}
\end{table}

\subsubsection{What social benefits are associated with fair practices, processes and technologies?}
\label{sec:se_rq3}
In the literature analyzed through the systematic review, three main perceived social benefits of fair systems were identified. The most frequently discussed benefit was \textit{general bias and discrimination mitigation}, which is consistent with the proposals for fairness definitions and metrics. The mitigation of discrimination is often linked to understandings of discrimination framed through distributive or procedural perspectives. The second most common benefit was \textit{trust and reliability}, showing that fairness is linked to increased user confidence. This is, of course, associated with discrimination mitigation, but also with factors such as auditability and transparency, which are sometimes linked to the overall fairness of a software system. Finally, \textit{legal compliance} was noted as an important benefit, indicating that fairness contributes to aligning systems with regulatory requirements and reducing legal risks, which may be explained by the influence of legal principles and frameworks that often guide fairness definitions.

\subsubsection{What approaches, methods, and practices are proposed to promote fairness in technologies?}
\label{sec:se_rq4}

The literature reports a diverse set of approaches, techniques, and procedures to promote fairness in systems. The most common approaches consist of \textit{techniques based on algorithm/model manipulation}, often categorized into pre-processing, in-processing, and post-processing, which address bias before model training, during training, or by adjusting model outputs through actions such as reweighting data, modifying decision boundaries, or applying fairness constraints. Complementing these, \textit{techniques based on data collection or manipulation} focus on improving the quality, representativeness, and diversity of datasets, including the construction of datasets with different demographic groups and the generation of synthetic data for underrepresented groups. \textit{Fairness toolkits} operationalize these technical approaches in a more plug-and-play style by providing built-in techniques aligned with different fairness objectives. Taking a less technical route, \textit{human-in-the-loop} approaches emphasize human participation as a fundamental element in the development of algorithmic systems to support ethical and fair practices, while also recognizing the limits of superficial implementations that may fail to address deeper socio-technical impacts. Connected to this, some studies discuss the importance of \textit{diversity in development teams}. Other practices focus on \textit{integrating fairness into the design process}, including \textit{fairness requirements elicitation}, stakeholder inclusion, and the \textit{literacy and education of both users and developers}. The literature also reports \textit{ethical and institutional frameworks}, including approaches aimed at tracing the causes and effects of discrimination within systems, as well as legal regulation, transparency, explainability, and auditing practices that operate at organizational and institutional levels.

\begin{table}[H]
\centering
\scriptsize
\caption{Software engineering articles per social benefits associated with fair practices, processes and systems}
\begin{tabularx}{\columnwidth}{|l|X|}
\hline
\textbf{Perceived benefit of a fair system} & \textbf{Article} \\
\hline
General bias/discrimination mitigation & P01, P02, P03, P06, P07, P09, P10, P11, P12, P14, P17, P18, P19, P21, P24, P25, P26, P27, P28, P29, P31, P37, P38, P41, P42, P43, P45 \\
\hline
Trust and reliability & P02, P05, P08, P15, P28, P30, P31, P34, P38, P41, P42, P43, P45 \\
\hline
Legal compliance & P02, P07, P08, P10, P11, P14, P34 \\
\hline
\end{tabularx}
\label{tab:se_rq3}
\end{table}

\begin{table}[!h]
\centering
\scriptsize
\caption{Software engineering articles per approach, technique and procedure for promoting fairness}
\label{tab:se_rq4}
\begin{tabularx}{\columnwidth}{|X|X|}
\hline
\textbf{Approaches, techniques and procedures} & \textbf{Articles} \\
\hline
Techniques based on algorithm/model manipulation & P01, P03, P04, P05, P06, P07, P08, P09, P10, P11, P12, P13, P14, P15, P16, P17, P19, P21, P22, P23, P24, P25, P26, P27, P28, P29, P31, P32, P33, P34, P35, P36, P37, P38, P40, P41, P42, P43, P44 \\
\hline
Techniques based on data collection or manipulation & P04, P05, P06, P07, P10, P11, P12, P13, P14, P16, P17, P19, P21, P22, P24, P25, P26, P27, P28, P29, P31, P32, P33, P34, P35, P36, P37, P38, P40, P41, P42, P43, P44 \\
\hline
Human-in-the-loop & P04, P10, P11, P12, P13, P23, P26, P33, P36, P39, P43, P44, P45 \\
\hline
Fairness toolkits & P04, P12, P13, P22, P27, P28, P29, P36, P40, P43, P44 \\
\hline
Institutional and ethical frameworks & P01, P02, P04, P12, P13, P23, P30, P38, P40 \\
\hline
Transparency and explainability & P15, P18, P30, P31, P38, P45 \\
\hline
Algorithmic legal regulation & P01, P18, P30, P38, P40 \\
\hline
Integrate fairness into the design process & P18, P36, P39, P44 \\
\hline
Auditing processes & P04, P18, P29 \\
\hline
Diversity in development team & P18, P22, P43 \\
\hline
Fairness requirements elicitation & P04, P17 \\
\hline
User/developer literacy and education & P18 \\
\hline
\end{tabularx}
\end{table}

\subsubsection{What limitations and challenges are identified in the efforts to achieve fairness in technologies?}
\label{sec:se_rq5}
The last question in this tertiary review focuses on the reported limitations and challenges of definitions and approaches for fairness. The most frequently mentioned challenge is the \textit{contextual aspect of fairness}, with studies arguing that fairness in machine learning cannot be understood or applied in isolation from the context in which decisions are made. A \textit{lack of consensus on fairness, bias, and their metrics} is also reported, as there is no universal agreement on what fairness means, how bias should be defined, or which metrics best capture these concepts, making it difficult to justify specific choices in practice. Several studies highlight the \textit{need for sensitive and vast data}, noting challenges related to privacy, consent, and security, and the resulting difficulty of implementing approaches that rely on the distribution of results according to sensitive attributes. The \textit{limitations of technical and statistical approaches} are also emphasized, as these approaches are often seen as insufficient to address the root causes of discrimination, even when contextual factors are acknowledged. The \textit{trade-off between fairness and performance} appears as a recurring challenge, including the \textit{impossibility to satisfy every metric}, conflicts between ethical goals and efficiency, and the \textit{computational complexity and resource usage} required by fairness techniques. Additional challenges include the \textit{limited interpretability and transparency of black-box models}, the \textit{complexity of complying with heterogeneous legal regulations}, and the \textit{insufficient consideration of intersectionalities} due to simplifications and false binaries in fairness approaches. Some studies also report \textit{developer bias}, which may be perceived as “objective” and reinforce discrimination under the guise of a “neutral” technology, as well as the \textit{abstract nature of fairness and difficulty of operationalization} through coding practices. Finally, a small number of studies mention the \textit{rapid pace of AI development}, which discourages careful fairness assessment, and the \textit{lack of studies in specific domains} as additional challenges.

\subsection{Humanities Systematic Literature Review}

We identified 25 studies in humanities-related domains that recently discussed fairness and fit the purposes of this study.

\subsubsection{How is fairness defined and conceptualized?}
\label{sec:hs_rq1}

The most common definition of fairness found in this study understands it as a \textit{sociopolitical and historically situated construct shaped by power and structural inequalities}. This perspective emphasizes that unfairness and discrimination are not isolated technical problems but are diffused through institutional infrastructures, governance policies, language, culture, and interpersonal dynamics, and are deeply connected to historical processes such as colonial power relations that continue to influence contemporary forms of discrimination, including algorithmic injustices. From this view, anchoring fairness solely in formal or distributional definitions is insufficient, as artificial intelligence systems are built within social structures that produce and maintain hierarchies and power relations. As a result, human science disciplines often shift the focus from asking whether an AI system is fair to asking how technologies reshape power and sociotechnical dynamics. A second definition, present in fewer studies, conceptualizes fairness as the \textit{outcome of legal anti-discrimination guidelines, institutional governance, and value-based frameworks}, where fairness is closely tied to legal standards, data protection, and governance practices that regulate the development and use of AI systems. The third conceptualization treats fairness as \textit{a normative and ethical principle}, in which moral reasoning guides judgments about algorithmic decisions by considering intentions, social effects, and ethical rules, as well as epistemic uncertainties and normative concerns arising from algorithmic decision making.

\begin{table}[!h]
\centering
\scriptsize
\caption{Humanities articles per fairness definition}
\label{tab:hs_rq1}
\begin{tabular}{|p{0.5\columnwidth}|p{0.4\columnwidth}|}
\hline
\textbf{Definition of fairness in human sciences disciplines} & \textbf{Articles} \\
\hline
A socio-political and historically situated construct shaped by power and structural inequalities. & P01, P02, P03, P04, P05, P07, P08, P09, P10, P11, P12, P13, P14, P15, P16, P17, P19, P20, P21, P22, P23, P24, P25 \\
\hline
The outcome of legal anti-discrimination guidelines, institutional governance, and value-based frameworks. & P05, P06, P08, P14, P17, P18 \\
\hline
A normative and ethical principle, with moral reasoning guiding judgments about the effects of algorithmic decisions. & P06, P09, P13 \\
\hline
\end{tabular}
\end{table}

\subsubsection{What social benefits are associated with fair practices, processes and technologies?}
\label{sec:hs_rq2}

The most frequently reported benefit consists in the \textit{reduction of structural injustices}, identified in all 25 analyzed articles. In contrast to the software engineering review, where discrimination reduction is often tied to outcomes, this literature emphasizes that reducing discrimination cannot be limited to the distribution of results produced by algorithms. An algorithm that complies with fairness metrics may still be considered unfair or unethical, as structural injustices are reproduced through technologies in ways that involve historical and colonial influences, institutional and legal actors, and broader relations of power and control. A second widely reported benefit is \textit{democratic participation and collective agency of technologies}, present in 19 articles, which emphasizes restoring agency and control to people and workers in contexts where automation and platform based systems have dispossessed them, often under neoliberal forms of governance. \textit{Equitable access and recognition}, discussed in 13 articles, is closely related to these benefits and focuses on the recognition of diverse forms of knowledge produced by minority groups that technologies may erase, raising questions about which forms of knowledge are prioritized and which are marginalized. Finally, \textit{disruption of entrenched power dynamics} refers to fairness as a means of challenging structural and normative foundations of inequality, where fair systems do more than distribute rights equally and instead transform social norms, redress harm, and build collective power among groups historically excluded from society and from technological systems.

\begin{table}[!h]
\centering
\scriptsize
\caption{Humanities articles per social benefits associated with fair practices, processes and systems}
\label{tab:hs_rq2}
\begin{tabularx}{\columnwidth}{|l|X|}
\hline
\textbf{Perceived benefit of a fair system} & \textbf{Article} \\
\hline
Reduction of structural injustices & P01--P25 \\
\hline
Democratic participation and collective agency over technologies & P01, P02, P03, P04, P10, P12, P13, P14, P16, P17, P18 P19, P20, P21, P22, P23, P24 \\
\hline
Equitable access and recognition & P01, P02, P04, P07, P13, P14, P15, P16, P17, P18, P19, P20, P21 \\
\hline
Disrupt entrenched power dynamics & P02, P04, P07, P09, P10, P14, P19, P20, P21, P22, P25 \\
\hline
\end{tabularx}
\end{table}

\subsubsection{What approaches, methods, and practices are proposed to promote fairness in technologies?}
\label{sec:hs_rq3}

The human and social sciences literature reports a set of approaches aimed at promoting fairness beyond technical interventions. \textit{Embedding multiple perspectives in system design}, present in 16 studies, seeks to prevent technologies from reproducing the narrow views of dominant groups by incorporating feminist, intersectional, and decolonial perspectives into design processes, often through interdisciplinary collaboration, to enable plural understandings of justice. Closely aligned with this, \textit{investigating the intersecting structures of inequality, oppression and exploitation}, reported in 8 articles, emphasizes analyzing how technologies reproduce colonial, racial, gendered, and capitalist power relations rather than treating bias as an isolated technical issue. \textit{Legal frameworks}, discussed in 10 studies, are viewed as mechanisms to translate ethical principles into enforceable norms, particularly through the combination of anti-discrimination law and data protection regulations, and are often connected to \textit{promoting transparency and accountability over technologies}, identified in 11 articles, which extends beyond explainable models to include visibility into institutional decision making, funding structures, and labor conditions. \textit{Participatory practices}, also present in 11 studies, focus on including those most affected by technologies in their design, evaluation, and governance, framing fairness as a collective and relational process that challenges corporate dominance. These approaches are supported by \textit{education and literacy about digital systems}, identified in 7 articles, which stress that fairness requires understanding not only how technologies work but also their social and ethical implications. Finally, \textit{ethical refusal}, found in 7 studies, argues that some technologies, such as surveillance systems, cannot be made fair at all, positioning fairness not only as a matter of design choices but also as a question of whether certain technologies should be developed in the first place.

\begin{table}[!h]
\centering
\scriptsize
\caption{Humanities articles per approach, technique and procedure proposed to promote fairness }
\label{tab:hs_rq3}
\begin{tabularx}{\columnwidth}{|X|X|}
\hline
\textbf{Approaches, techniques and procedures} & \textbf{Articles} \\
\hline
Embedding multiple perspectives in system design & P01, P02, P07, P09, P10, P11, P12, P14, P15, P16, P19, P20, P21, P23, P24, P25 \\
\hline
Participatory practices & P02, P03, P04, P07, P13, P14, P15, P19, P21, P22, P24 \\
\hline
Legal frameworks & P01, P02, P04, P05, P06, P08, P09, P14, P17, P18 \\
\hline
Investigating the intersecting structures of inequality, oppression and exploitation & P02, P07, P16, P19, P20, P21, P22, P25 \\
\hline
Promoting transparency and accountability over technologies & P02, P08, P09, P12, P16, P18, P21, P22 \\
\hline
Education and literacy about digital systems & P01, P02, P10, P12, P13, P14, P19 \\
\hline
Ethical refusal & P02, P03, P13, P15, P19, P20, P22 \\
\hline
\end{tabularx}
\end{table}

\subsubsection{What limitations and challenges are identified in the efforts to achieve fairness?}
\label{sec:hs_rq4}

The main challenge identified concerns the \textit{failure of AI fairness efforts to recognize and address the complexity of social life}. Present in 20 articles, this critique shows how AI practitioners often treat sociotechnical problems as technical and formal issues, reducing the complexity of lived material experiences into quantifiable categories and treating fairness interventions as mere fixes without questioning why bias exists or how it is perpetuated and complexified by technologies. Closely related to this, \textit{reproduce hierarchical and historical structures of discrimination and power}, identified in 10 articles, argues that AI fairness practices often replicate the same structural inequalities and power relations they claim to address, mirroring liberal anti-discrimination approaches that individualize injustice, ignore structural causes, and protect existing hierarchies of race, gender, and capital. \textit{Non-transparency and accountability gaps}, discussed in 13 articles, further reinforce this reproduction when harms cannot be attributed to responsible actors and structural discrimination is framed as a technical failure, requiring responses that go beyond explainable AI toward institutional transparency and participatory governance. These gaps are intensified by the \textit{lack of comprehensive regulation}, reported in 3 studies, which note that existing legal frameworks are insufficient to address sociotechnical complexities and often prioritize innovation and profit through corporate self-governance and voluntary ethical codes, while failing to account for broader social dynamics, civil liberties, and privacy concerns.

\begin{table}[!h]
\centering
\scriptsize
\caption{Humanities articles per limitations and challenges are identified in the efforts to achieve fairness}
\label{tab:hs_rq4}
\begin{tabularx}{\columnwidth}{|X|X|}
\hline
\textbf{Perceived benefit of a fair system} & \textbf{Article} \\
\hline
Failure of AI-practitioner efforts to recognize and address the complexity of sociotechnical dynamics & P01, P02, P03, P04, P06, P07, P10, P12, P13, P14, P15, P16, P17, P19, P20, P21, P22, P23, P24, P25 \\
\hline
Technical fairness practices reproduces hierarchical structures of discrimination and power & P01, P02, P13, P16, P19, P20, P21, P22, P24, P25 \\
\hline
Non-transparency and acountability gaps & P06, P07, P08, P09, P10, P11, P12, P14, P19, P20, P21, P22, 24, P25 \\
\hline
Lack of comprehensive regulation & P08, P14, P25 \\
\hline
\end{tabularx}
\end{table}

\subsection{Cross-Domain Integration: Software Engineering and Humanities Perspectives}
Across both domains, the findings reveal that software engineering and the humanities approach fairness through fundamentally different yet potentially complementary lenses. Software engineering research defines fairness as a measurable property of systems, emphasizing formal metrics, verifiability, and performance trade offs to ensure equitable outcomes. In contrast, humanities scholarship frames fairness as a sociopolitical and historically situated construct rooted in power, inequality, and justice. While the engineering perspective operationalizes fairness into quantifiable procedures, the humanities perspective shows how such formalization can obscure the structural and historical causes of discrimination. Overall, these perspectives point to both the necessity and the limitation of reducing fairness to technical optimization, indicating that fairness in software systems depends on connecting measurable criteria with contextual and ethical understanding that situates technology within broader social realities. Table \ref{tab:conceptual_characterization} summarizes our analysis. 

\begin{table}[h!]
\centering
\scriptsize
\caption{Conceptual characterization of fairness across domains}
\label{tab:conceptual_characterization}
\begin{tabular}{|p{1.3cm}|p{1.5cm}|p{4.5cm}|}
\hline
\textbf{Domain} & \textbf{Conceptual Orientation} & \textbf{Characterization of Fairness} \\ \hline

\textbf{Software Engineering} &
Technical, measurable, and algorithmic perspective &
\textit{Metric-oriented, system-focused, verification-driven, model-centric, performance-aware, data-dependent, and compliance-focused.}
Fairness is treated as a quantifiable quality attribute integrated into system design, measurable through formal metrics, and bounded by performance and compliance constraints. \\ \hline

\textbf{Human Sciences} &
Ethical, contextual, and justice-oriented perspective &
\textit{Justice-oriented, context-aware, power-focused, value-driven, participatory, intersectional, and critical-reflective.}
Fairness is framed as a historically, ethically, and politically situated value that demands contextual, participatory, and critical engagement. \\ \hline

\textbf{Intersection} &
Socio-technical, responsible, and integrated perspective &
\textit{Accountability-focused, transparency-driven, trust-enhancing, diversity-aware, education-centered, ethics-integrated, and contextually measurable.}
Fairness is conceptualized as a co-constructed socio-technical practice linking systems and society, combining measurable rigor with contextual and ethical awareness. \\ \hline

\end{tabular}
\end{table}

\section{Discussion}
\label{sec:discussion}
In this section we answer our research questions, compare our findings with the literature and discuss the implications of our findings.

\subsection{Answering the Research Questions}

Our first RQ was {\textit{How is fairness defined and conceptualized?}}
In software engineering, fairness is defined as a measurable property of systems, emphasizing balanced outcomes and verifiable performance. In the humanities, fairness is viewed as a social and political construct rooted in justice and power. The first treats fairness as technical optimization; the second as contextual transformation. The second RQ was {\textit{What metrics and criteria are used to evaluate fairness?}}
In software engineering, fairness is evaluated through statistical measures such as demographic parity and equal opportunity. These metrics help detect bias but confine fairness to what can be quantified, overlooking broader social realities. Our third RQ asked {\textit{What social benefits are associated with fair practices, processes, and technologies?}}
Both domains link fairness to reducing discrimination and building trust. Engineering emphasizes reliability, transparency, and compliance, while the humanities focus on justice, equity, and empowerment. The fourth RQ was {\textit{What approaches, methods, and practices are proposed to promote fairness in technologies?}}
Engineering promotes algorithmic adjustments, fairness toolkits, and human-in-the-loop design. The humanities emphasize participatory design, interdisciplinary collaboration, and ethical reflection on whether technologies should exist at all. Finally, the fifth RQ was {\textit{What limitations and challenges are identified in the efforts to achieve fairness in technologies?}}
Engineering faces challenges with data quality, metric conflicts, and performance trade-offs. The humanities highlight structural inequities, weak regulation, and the reduction of social problems to technical ones. Both agree that fairness requires attention to context, ethics, and human impact.

\subsection{Comparing and Contrasting Findings}

After exploring fairness in both software engineering and the human sciences, we discuss the limits of engineering approaches to fairness to expand these definitions and inspire new perspectives for building fair systems. Fairness in software engineering is tied to decision-making systems that classify and organize the world. A classification system assumes completeness and total coverage of what it describes (\cite{bowker_star_sorting_2008}), but such completeness is impossible in sociotechnical contexts. A system may appear “fair” for assigning result A to person B, yet that fairness cannot be assumed across different times, contexts, and institutional realities. Reducing fairness to the distribution of results hides the situated and material conditions that shape those outcomes, which raises the question of what forms of sociotechnical interaction become invisible through such simplification.

This limitation becomes clear when examining real-world cases. In Rio de Janeiro, several Uber drivers were killed after navigation algorithms sent them through areas controlled by militias \cite{g1_2024, oglobo_2024, cnn_2024}. This situation cannot be explained through fairness as distribution of results, since no metric captures how an algorithm interacts with local violence and geography. Algorithms shape urban experience without accounting for it, overlooking sociopolitical conditions such as public safety. As \cite{hoffmann_where_2019, west_redistribution_2020} observe, AI practitioners often fail to consider what it truly means to design fair technologies, reproducing existing power structures and discrimination. A fairness logic centered only on balanced results therefore obscures the social realities within which technologies operate.

Procedural fairness appears more inclusive but remains confined to mathematical logic. As \cite{pessach_review_2023} notes, assessing fairness depends on context and perception, yet treating context as external assumes that society exists apart from engineering. In practice, process becomes a statistical mechanism that seeks fair features and balances accuracy while ignoring the social consequences of those choices. This view risks absolving algorithms of responsibility by claiming fairness when sensitive attributes are removed (\cite{gonzalez-sendino_review_2024}), which is the baseline for fairness through unawareness. However, as \cite{schwobel_long_2022} argues, sensitive data such as race or gender relate to other attributes like education or income, and ignoring them conceals rather than removes bias, reinforcing existing inequalities.

These fairness conceptions focus mainly on discrimination yet fail to address the complex sociotechnical interactions that generate it. Even within their technical scope, they do not produce systems free from discrimination (\cite{hoffmann_where_2019, greene_better_2019, weinberg_rethinking_2022}). \cite{xue_bias_2023} and \cite{papagiannidis_responsible_2025} highlight the need to engage actors beyond organizational boundaries, while \cite{santos_software_2024} identifies design and historical biases as sources of fairness debts. However, as long as fairness is framed through cause and effect models, algorithms simply carry forward existing inequities. Following \cite{latour2005reassembling}, replacing causes with interacting actors offers a more material understanding of fairness as a collective and contextual process rather than a technical adjustment.

Formal definitions of fairness extend these limits by reducing fairness to a verifiable system property. Although algorithms “speak mathematics” (\cite{rabonato_systematic_2025}), they are more than sequences of logic and computation. As \cite{pham_fairness_2023} observes, embedding fairness within the development process allows auditing and verification but disconnects fairness from the material realities in which technologies operate. Software engineers define fairness through what they can formalize, stripping away ethical and political meaning in favor of operational clarity. Fairness in this context serves the logic of a system that values only what can be codified and tested, leaving out the social and moral dimensions that make technologies just or unjust.

Ultimately, defining fairness as a formal, procedural, or distributive property reveals how computer science knowledge practices operate as practices of absence (\cite{malazita_infrastructures_2019}), excluding the complexities of human experience in favor of measurable criteria. Metrics such as demographic parity and conditional statistical parity (\cite{balayn_managing_2021, kheya_pursuit_2024}) reduce lived experiences to variables and disregard the structural conditions that precede algorithmic decisions. Sensitive attributes like race, gender, and class are not simply data points but markers of material life that algorithms erase. As shown in university admissions systems, such simplifications can reproduce inequality rather than remedy it. Scholars from the humanities therefore call for ethical refusal, questioning whether some technologies should be built at all (\cite{weinberg_rethinking_2022, stark_breaking_2023}), and for a structural view of fairness that confronts the hierarchies and systems of power technologies often reproduce (\cite{zajko_conservative_2021}).

\subsection{Implications for Research and Practice}

For research, our results indicate the need to move beyond viewing fairness as a fixed or measurable system property. Existing models often rely on formal and procedural definitions centered on classification and distribution, despite the fact that sociotechnical contexts are dynamic and incomplete. Future research should therefore focus on how fairness is produced within networks of people, data, and institutions, paying attention to the relationships between technical mechanisms and the material, social, and historical conditions in which technologies are developed and used. Approaches that combine quantitative evaluation with qualitative and participatory inquiry can support more interdisciplinary and context-grounded fairness research. For practice, our findings indicate that fairness cannot be achieved solely through audits, metrics, or formal verification. Developers and organizations need to account for how algorithmic decisions interact with users’ lived realities, including geography, safety, and inequality. Treating fairness as a purely mathematical concern obscures the social and political dimensions of technology. Fairness should instead be approached as an ongoing process involving negotiation, accountability, and reflection, which includes engaging diverse communities, documenting trade-offs transparently, and recognizing the ethical consequences of technical systems. In some cases, achieving fairness may also require refusing to design or deploy technologies that perpetuate harm.
\section{Conclusions and Future Work} \label{sec:conclusions}

This study compared how fairness is understood and practiced in software engineering and in the human sciences. The analysis of seventy studies revealed a divide between the formal, metric-based definitions prevalent in software engineering and the socially situated interpretations emphasized in the humanities. While software engineering often treats fairness as a measurable attribute aimed at bias mitigation and accountability, the human sciences frame it as a political and relational concept tied to power and social structures. Bridging these perspectives requires rethinking fairness not only as a system property but also as a situated practice shaped by context and values. Our future work will extend this analysis by collecting the experiences of software practitioners, enabling a tripartite comparison between software engineering literature, software engineering practice, and the social sciences. This integration aims to advance fairness as an ethical principle guiding the design and governance of sociotechnical systems.

\section{Data Availability}

The papers analyzed in this scoping study are available at \url{https://figshare.com/s/83d6c294f2f6e4b2630d}

\nocite{*}

\bibliographystyle{ACM-Reference-Format}
\bibliography{bibliography}

\appendix

\end{document}